\documentclass[journal]{IEEEtran}

\usepackage[dvips]{graphicx}

\usepackage{amsmath}
\usepackage[cmintegrals]{newtxmath}

\usepackage{amsmath}
\usepackage{amsfonts,amssymb}
\usepackage{multirow}
\usepackage{color}
\usepackage{scalerel}

\def\bF{{\mathbb{F}}}

\def\argmax{\mathbb{\rm argmax}}

\def\GEO{\mathop{\rm GEO}}
\def\MEO{\mathop{\rm MEO}}
\def\LEO{\mathop{\rm LEO}}

\def\soft{\mathop{\rm soft}}
\def\hard{\mathop{\rm hard}}
\def\TW{\mathop{\rm TW}}
\def\OW{\mathop{\rm OW}}

\begin{document}
\title{Two-Way Physical Layer Security Protocol for Gaussian Channels
\thanks{MH was supported in part by JSPS Grant-in-Aid for Scientific Research (A) No.17H01280 and for Scientific Research (B) No.16KT0017, and Kayamori Foundation of Informational Science Advancement.
}}

\author{Masahito~Hayashi,~\IEEEmembership{Fellow,~IEEE} 
        and
\'{A}ngeles V\'{a}zquez-Castro,~\IEEEmembership{Senior Member,~IEEE,}
\thanks{Masahito Hayashi is with
Graduate School of Mathematics, Nagoya University, Nagoya, Japan,
Shenzhen Institute for Quantum Science and Engineering, Southern University of Science and Technology, Shenzhen, China,
Center for Quantum Computing, Peng Cheng Laboratory, Shenzhen, China
and
Centre for Quantum Technologies, National University of Singapore, Singapore.
e-mail: masahito@math.nagoya-u.ac.jp}
\thanks{
\'{A}ngeles Vazquez-Castro
is with the Department of Telecommunications and Systems Engineering, and with The Centre for Space Research (CERES) of Institut d'Estudis Espacials de Catalunya (IEEC-UAB) at 
Autonomous University of Barcelona,
Barcelona, Spain
e-mail: angeles.vazquez@uab.es.}
\thanks{Manuscript submitted 30th June 2018; revised xxx, 2017.}}

\maketitle

\begin{abstract}
In this paper we propose a two-way protocol of physical layer security using the method of privacy amplification against eavesdroppers. 
First we justify our proposed protocol by analyzing the physical layer security provided by the classic wiretap channel model (i.e. one-way protocol). 
In the Gaussian channels, the classic one-way protocol requires Eve's channel to be degraded w.r.t. Bob's channel. 
However, this channel degradation condition depends on Eve's location and whether Eve's receiving antenna is more powerful than Bob's.
To overcome this limitation, we introduce a two-way protocol inspired in IEEE TIT (1993)  that eliminates 
the channel degradation condition.

In the proposed two-way protocol, on a first phase, via Gaussian channel,
Bob sends randomness to Alice, which is partially leaked to Eve. 
Then, on a second phase, Alice transmits information to Bob over a public noiseless channel. 
We derive the secrecy capacity of the two-way protocol when 
the channel to Eve is also Gaussian.
We show that the capacity of the two-way protocol is always positive. 
We present numerical values of the capacities illustrating the gains obtained by our proposed protocol. 
We apply our result to simple yet realistic models of satellite communication channels.
\end{abstract}


\section{Introduction}
Physical layer security for wireless communications has become a major research topic in recent years because it does not need the computational assumption \cite{InfoTheoreticSec,PhysicalLayerSecurity,Wu2018}. 
Different properties of the wireless channel can be exploited using information theoretical tools to prevent leakage of information towards potential eavesdroppers. The classic wiretap model as first proposed by Wyner \cite{Wyner1975} and then generalised by I.~Csisz\'ar and J.~K\"orner \cite{Csiszar1978} was later strengthened to meet cryptographic security standards in~\cite{Csiszar1996} and \cite{hay-wire}, 
the latter framed within spectrum information-theoretic methods \cite{Han}. 
We adopt such approach here: we assume the physical layer security realized by 
a stochastic wiretap encoder \cite{Hayashi2011,H13,Bloch2015} based on the privacy amplification method \cite{BBCM,HILL}. 
This method decouples reliability and secrecy, enabling the implementation of different security protocols. 

\begin{figure}[tbh]
\centering
\includegraphics[scale=0.3]{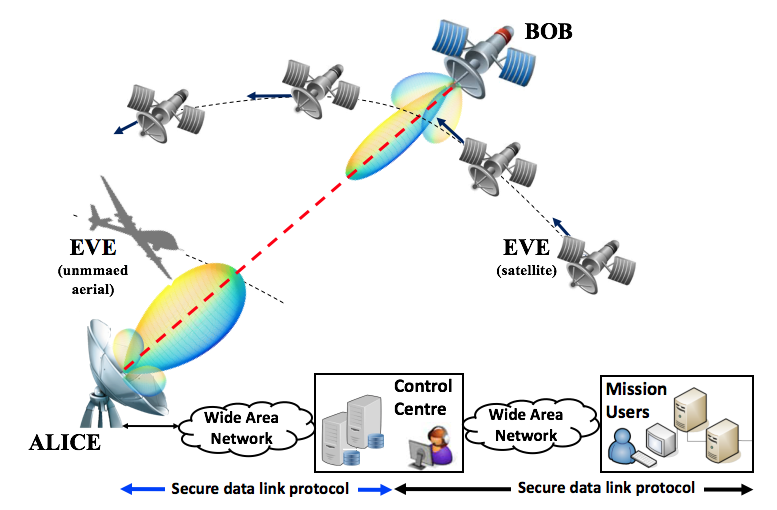}
\protect\caption{Illustration for our satellite communication scenario.}
\label{ICC}
\end{figure}

In its simplest implementation, the wiretap channel model using privacy amplification can be realized as a one-way security protocol whereby Alice sends a keyless secret message to Bob protected with universal$_2$ hash functions \cite{Wegman1981,Krawczyk,Hayashi-Tsurumaru}. 
In the Gaussian wiretap channel,
secrecy capacity is positive as long as 
Eve's channel has a worse signal to noise ratio than the channel between Alice and Bob, 
i.e., Eve's channel is degraded w.r.t the main channel between Alice and Bob \cite{Hellman1978}. 
The same holds to ensure positive secrecy rate in the finite-length \cite{PhySecSat,arXiv}. 
However, 
when the channel between Alice and Bob has a worse signal to noise ratio than the channel to Eve,
we cannot realize secure communication in this scenario \cite{Liang2006,GLG,EU}.
For example, in the satellite communication,
Eve's satellite usually stays in lower orbit than the orbit of Bob's satellite like in the example scenario in Fig. \ref{ICC}, 
which implies that Eve has better signal to noise ratio than Bob.
That is, it is hard to realize secure satellite communication 
with the above proposals of wiretap codes 
if we cannot identify Eve's spatial locations.
This type of attack is often called passive man-in-the-middle attack \cite{Conti2016}.

To resolve this problem, 
the papers \cite{WGH,FJHWY,ZWPT,TY} introduced two-way protocols, in which,
the channels of both directions are noisy Gaussian channels.
However, when both channels are noisy Gaussian channels,
there still exists a possibility that we cannot realize secure communication 
dependently of Eve's and Bob's spatial locations.
To overcome this limitation, Maurer \cite{Maurer1993} proposed 
a two-way protocol based on the binary symmetric channel.
In this paper,  inspired by Maurer's idea,
we propose a two-way protocol with Gaussian wiretap channel and 
public noiseless feedback, in which,
the feedback channel is given as a public noiseless channel with discrete variable.
This paper assumes that the noise in 
the channel of the initial transmission from Bob to Alice
is independent of the noise in 
that to Eve while the paper \cite{Correlation} considers the case when these two noises are correlated.
Under this assumption, unless 
the channel of the initial transmission to Eve
is noiseless,
this protocol always has positive secure transmission rate regardless Eve's and Bob's spatial locations.
In particular, we focus on the simple but realistic (for fixed user terminals) Gaussian-channel satellite system scenarios. 
Our results demonstrate that our two-way protocol greatly outperforms 
state-of-the-art one-way and two-way protocols because our two-way protocol always realize secure communication under passive man-in-the-middle attack independently of Eve's spatial location.
Note that extension of our results to channels with fading is straightforward, but we take this purely averaging calculation problem out of the scope of our paper, here we focus on the protocols for the Gaussian channel
without fading. More specifically, on the application to the Gaussian with BPSK satellite channel, which is considered in current satellite communication standards \cite{DVB-S2X}
Further, we can equip authentication in our protocol
by attaching universal$_2$ hash function \cite{Wegman1981,Krawczyk}.
This protocol can prevent active man-in-the-middle attack.
Therefore, the advantages of our protocol are summarized as follows.

Contribution 1) We address wiretap channel security problem (i.e. eavesdropping) and propose for the first time a novel practical Gaussian wiretap protocol implementing theoretical ideas in Maurer's paper \cite{Maurer1993}.

Contribution 2) Our novel two-way protocol greatly outperforms state-of-the-art one-way protocol because our protocol shows always positive secrecy capacity independently of Eve's location (i.e. it does not require channel degradation condition of one-way wiretap channel). This is our main technical result based on novel and rigorous information theoretical proof.

Contribution 3) Our practical two-way protocol outperforms other proposals of two-way protocols \cite{WGH,FJHWY,ZWPT} in the following sense. 
First, because while other two-way protocols require several communication rounds, our protocol only requires two rounds. Second, because other two-way protocols while outperforming one-way protocol, they still may have negative secrecy capacity. Finally, because our protocol only requires two rounds, our protocol is highly
suitable to secure communication channels with large delay, e.g. satellite channels.

Contribution 4) We show the performance of our protocol with meaningful numerical results for the realistic Gaussian BPSK modulated satellite channel, which is included in current satellite communication standards \cite{DVB-S2X}. For this, we use our system modeling which allows to evaluate the security capacity as a function of the system parameters. Hence, our method and results are useful for secure communication design.

One might consider that the real feedback channel is also a noisy channel.
However, if we choose sufficiently strong intensity and a
suitable error correcting code, the information transmission of
the feedback channel can be regarded as a noiseless channel.
In this case, we can regard the feedback channel as a noiseless
public channel with discrete variable. 
In contrast, the noise of
the initial Gaussian channel is essential because its presence
makes the difference between mutual informations from Alice
to Bob and Eve. 
Furthermore, the information leakage in the channel during the second phase needs not be considered because information leakage is only relevant on the first phase.
Hence, it is allowed to make the power of the transmission signal very strong
in the second phase
while we cannot use such a strong power in the first phase
to control the secrecy.
Since the information transmission rate with the strong power 
is much larger than that with the weak power,
the consuming time of the first phase is dominant
in comparison with that of the second phase is dominant.
That is, the first phase is the bottleneck in this setting.
Therefore, this paper optimizes the amount of
noise in the initial Gaussian channel to maximize the wiretap
capacity in this model.

In fact, the paper \cite{RealisticChannel} considers a similar topic.
It is a follow up of this submission with practical focus.
Hence, it contains only the brief description of the two-way protocol without proper proof.
Also, the analysis of the secure satellite communication in \cite{RealisticChannel} is different from our analysis in Sections IV and V.
That is, it did not consider the optimization while the analysis in this paper is based on the optimization given in Section IV.

Relations to other studies are summarized as follows.
While the paper \cite{TY} discusses two-way wiretap channels,
it considers a new scheme cooperative jamming.
The scheme in \cite{TY} has several users that are cooperative
while this paper has only two cooperative users, the legitimate sender and the legitimate receiver.
Therefore, the method in this paper cannot be compared with \cite{TY}.
In the paper \cite{AFJK}, Bob feeds back some randomness that is used as a secret key,
exactly like the present manuscript. 
However, it assumes that the feedback is noiseless and
secure, so Eve does not observe it, which is different from here.
In the paper \cite{BPS},
unlike the present work, Bob does not control the feedback link. 
However, it allows all players to observe everything.
Hence, the method in this paper cannot be compared with 
the results \cite{AFJK,BPS}.

Our work is structured as follows. In Section II, 
we review the results of one-way standard protocol. 
In Section III, we propose our two-way protocol.
In Section IV, we make numerical optimization for our obtained secret capacities.
In Section V, we  apply our result to simple realistic models of satellite communication.
Finally in Section VI we discuss the protocol and draw some conclusions.
\section{One-Way Physical Layer Security Protocols}
\subsection{Signal and channel Model}
First, we review the results of one-way standard protocol with Gaussian channels, which model wireless channels in relevant realistic communication scenarios such as satellite radiofrequency communication channels. 
The signal variables received by Bob and Eve
can be modeled as
\begin{equation}\label{eq:SysModel-P}
Y'=\sqrt{E_s}V +\sqrt{n_{B}} N_1, \quad
Z'=\gamma_g\sqrt{E_s}V +\sqrt{\gamma_n n_{B}} N_2,
\end{equation}
where $V$ is a variable modeling the transmitted signal and $E_s$ is the energy per symbol expressed in Joules. 
$Y'$ is the random variable representing the signal received at the legitimate receiver, $Z'$ the random variable representing the signal received at the eavesdropper's receiver and $N_1$ and $N_2$ are zero-mean circular complex Gaussian random variables with unit variance. $n_B$ is the noise spectral density power of Bob's receiver expressed in Joules per Hertz. The coefficient $\gamma_g$ models the amplitude attenuation of the wiretapper's channel w.r.t. the legitimate channel. The analytical expression of $\gamma_g$ will depend on the system under analysis as well as the channel assumptions and corresponding time scale. The multiplicative coefficient $\gamma_n$ expresses wiretapper's receiver noise with respect to Bob's receiver noise. Denote the signal-to-noise ratio (SNR) for Bob as 
$$ \eta_B = \frac{E_s}{n_{B}}=\frac{P}{N_{B}}$$
where $P$ and $N_B$ are the system and noise power at Bob's receiver, respectively both expressed in Watts. We can then rewrite the signal model as
\begin{equation}\label{eq:SysModel}
Y=\sqrt{\eta_B}V+N_1, \quad
Z=\gamma_B\sqrt{\eta_B}V+N_2;
\end{equation}
where $Y = Y'/\sqrt{n_{B}}$, $Z = Z'/\sqrt{\gamma_n n_{B}}$ and $\gamma_B := \gamma_g/\sqrt{\gamma_n}$. 
Hence, $N_1$ and $N_2$ are 
zero-mean circular complex Gaussian random variables with unit variance. 

\begin{figure}[tbh]
\centering
\includegraphics[scale=0.3]{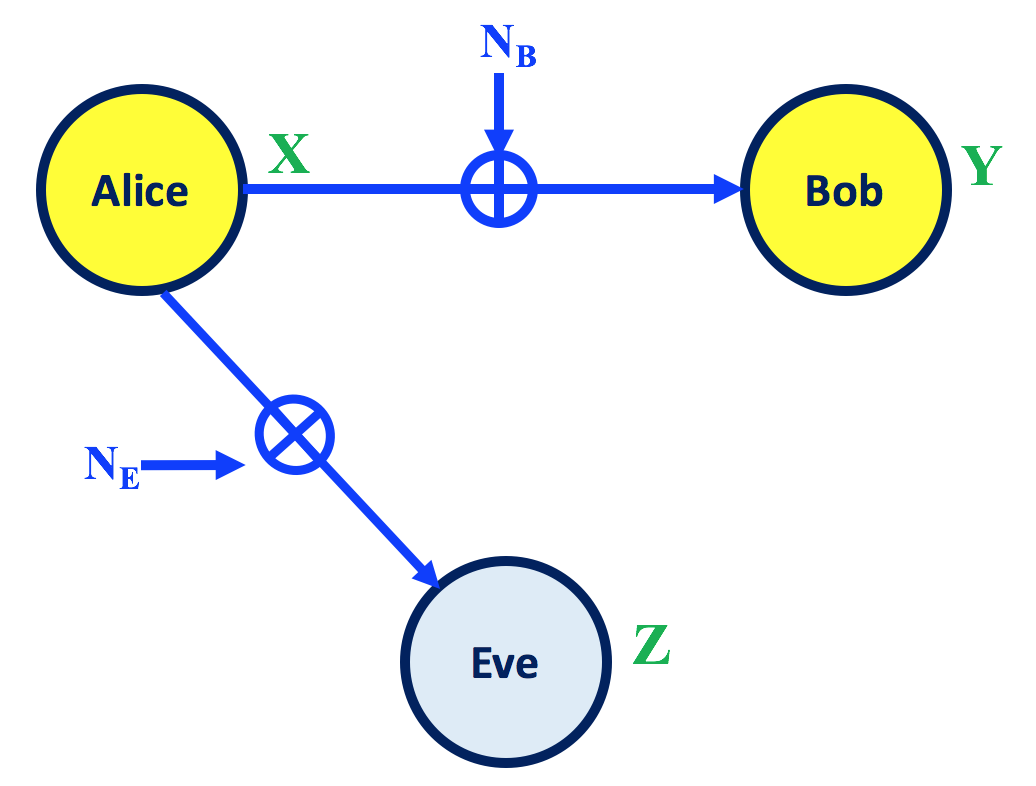}
\protect\caption{Graphical illustration of the one way protocol with Gaussian channels.
$N_B$ and $N_E$ are noise powers at Bob's and Eve's receivers.
}
\label{fig:OneWayProtocol}
\end{figure}

\subsection{Secrecy capacity with BPSK modulation and soft decision}
In the one-way model, Alice sends the encoded information to Bob via channel  \eqref{eq:SysModel} as
Fig. \ref{fig:OneWayProtocol}.
Now, we assume the BPSK modulation, in which, 
Alice encodes her binary information $A \in \bF_2$ to $V=(-1)^A$.
In this scenario, 
the secrecy capacity $C^{\OW}_{\soft}(\gamma_B,\eta_B)$
for the one-way protocol is given as \cite[(46)]{arXiv}
\begin{align*}
C^{\OW}_{\soft}(\gamma_B,\eta_B)
=&I (A; Y ) - I (A; Z) \nonumber \\
=&\int_{-\infty}^\infty 
u \Bigg[ \frac{1}{\sqrt{8 \pi}}\Big(e^{-\frac{(y-\sqrt{\eta_B} )^2}{2}}
+e^{-\frac{(y+\sqrt{\eta_B} )^2}{2}}\Big)\Bigg]
d y \nonumber \\
&-\int_{-\infty}^\infty 
u \Bigg[ \frac{1}{\sqrt{8 \pi}}\Big(
e^{-\frac{(z-\gamma_B\sqrt{\eta_B} )^2}{2}}
\!+\!e^{-\frac{(z+\gamma_B\sqrt{\eta_B} )^2}{2}}
\Big) \Bigg]
d z ,
\end{align*}
where $u(x):= - x \log x$
when 
\begin{align}
\gamma_B<1 , \quad {\rm i.e.,} \quad \gamma_g^2 <\gamma_n.
\label{LPK}
\end{align}
Also, when the condition (\ref{LPK}) does not hold, 
the capacity is zero. In this case, we cannot realize secure communication in this scheme.
Here,
Bob and Eve are assumed to store the sequence of the received continuous signals
and apply the decoder to them. 
This type of information processing is called soft decision decoding \cite{Proakis2001}.

\subsection{Secrecy capacity with BPSK modulation and hard decision}
To save the cost of decoding, 
the receiver converts the received continuous signal to a binary signal in the reception
and apply the decoder to the sequence of the binary signals.
This type of information processing is called hard decision decoding \cite{Proakis2001}.
When the receiver applies this method, 
it is sufficient to store only binary signals, which saves the memory of the receiver.
Here, as another scenario, we consider the case when Bob and Eve obtain $B$ and $E$ using hard decision detection on their received signals $Y$ and $Z$ as defined in the previous sections, respectively. 
The crossover probability between Alice and Bob induced by the Bernoulli random variable $X_1$ is given as $\epsilon_B^*= 0.5 \mbox{erfc}(\sqrt{\eta_B/2})$ and the crossover probability between Alice and Eve induced by the Bernoulli random variable $X_2$ is given as $\epsilon_E^*= 0.5 \mbox{erfc}(\gamma_B \sqrt{\eta_B/2})$ with
\begin{align}
\mbox{erfc}(t):=& \frac{2}{\sqrt{\pi}} \int_{t}^{\infty}e^{-t^2}dt  .\label{LBH}
\end{align}
In this scenario, 
by using the binary entropy function $h(x):= -x \log x -(1-x)\log (1-x)$,
the secrecy capacity 
for the one-way protocol is given as 
\begin{align}
C_{\hard}^{\OW}(\gamma_B,\eta_B)
= h(\epsilon_E^*)-h( \epsilon_B^*),
\end{align}
which is positive only when $\epsilon_E^* > \epsilon_B^*$, which is equivalent to \eqref{LPK}.

\section{Two-Way Physical Layer Security Protocol with Gaussian channels and BPSK modulation}

\subsection{Signal and channel Model}
One-way model requires the condition that 
the mutual information between the sender and the legitimate receiver 
is larger than that between the sender and the eavesdropper.
This assumption does not hold when 
the eavesdropper performs passive man-in-the-middle attack.
To resolve this problem, we consider two-way protocol for the Gaussian channel and BPSK modulation 
as follows. 
In an initial step, Bob sends a random variable $V$ to Alice.
In this case, Alice and Eve obtain the variables $Y$ and $Z$, respectively, as follows.
\begin{align}
Y=\sqrt{\eta_A}V+N_1, \quad
Z=\gamma_A\sqrt{\eta_A}V
+N_2,
\label{eq:GaussChannel}
\end{align}
where $N_1$ and $N_2$ are zero-mean circular complex Gaussian random variables with unit variance and the coefficient $\gamma_A$ models the amplitude attenuation of the wiretapper's channel w.r.t. the legitimate channel (which is now between Bob and Alice and not between Alice and Bob). 
Notice that the transmitter of the noisy Gaussian channel \eqref{eq:GaussChannel} is not Alice who is
the sender of the secret message of this protocol.
Therefore $\eta_A$ is now 
$$ \eta_A = \frac{E_s}{n_{A}}=\frac{P}{N_{A}},$$
where $n_A$ is the noise spectral density power of Alice's receiver expressed in Joules per Hertz. 

\begin{figure}[tbh]
\centering
\includegraphics[scale=0.4]{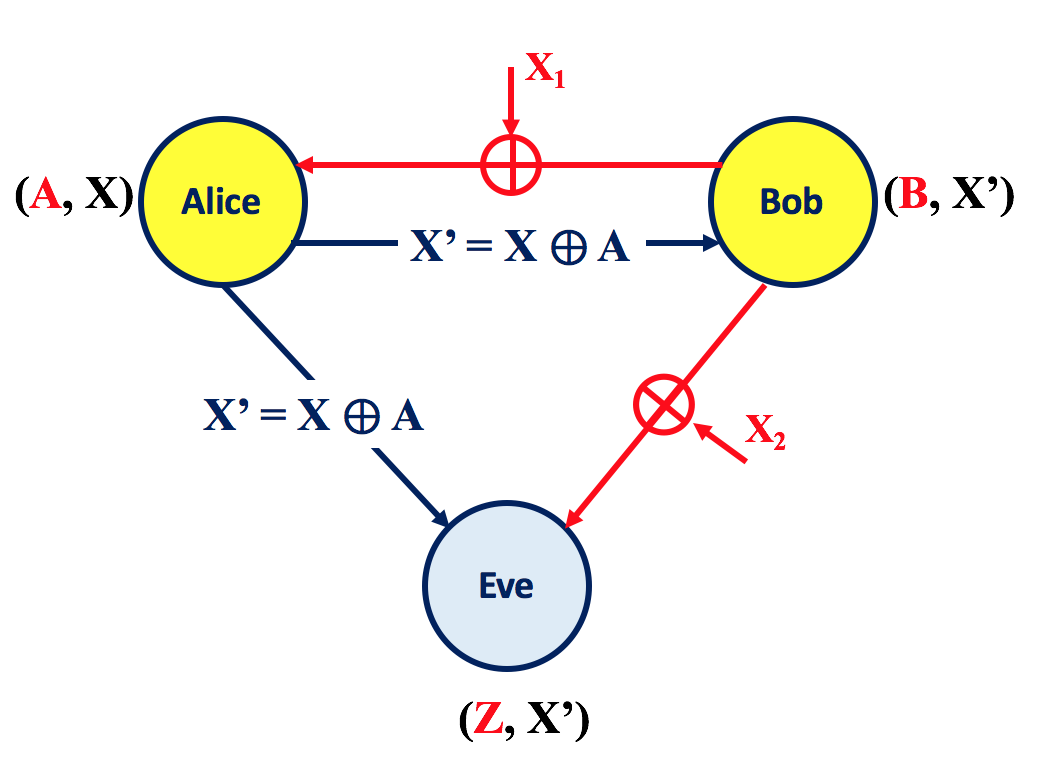}
\protect\caption{Graphical illustration of the two-way protocol with Gaussian channels. 
Phase 1 is shown in red and Phase 2 is shown in black.}
\label{fig:TwoWay_Gaussian}
\end{figure}

Bob generates the binary variable $B \in \bF_2$ subject to the binary uniform distribution, and sends $V = (-1)^{B}$ via the above RF channel. Applying hard decision decoding to $Y$, Alice obtains the binary variable $A$. 
In the next step, Alice prepares another binary variable $X$, and sends $X':= X\oplus A$ to Bob via a public channel. When $X$ is regarded as the channel input information, the legitimate receiver's output is $B$ and $X'$ while the eavesdropper's output is $Z$ and $X'$. The overall process along with the generated random variables is shown in Fig. \ref{fig:TwoWay_Gaussian}.

\subsection{Protocol}
Based on the above discussion, we fully describe our concrete protocol.
For this aim, we fix an error correction code, i.e., the pair of the encoder $\phi_{e,n}$ and 
the decoder $\phi_{d,n}$ with block length $n$. 
Then, 
combining the error correction code and universal$_2$ hash functions \cite{Wegman1981,Krawczyk,Hayashi-Tsurumaru},
we employ the wiretap code given in 
\cite[Appendix A]{arXiv} 
using random seed $S$ and we have the wiretap encoder $\phi_{e,n|S}$ and the wiretap decoder $\phi_{d,n|S}.$ 
This code construction achieves the strong secrecy even in the continuous system \cite{Hayashi2011,PhySecSat,arXiv}.

Then, we propose the following two way protocol.
\begin{description}
\item[(1)]
Bob generates the binary data sequence $B_1, \ldots, B_n \in \bF_2$.
Then, sends $(-1)^{B_1}, \ldots, (-1)^{B_n}$ via the channel described in (\ref{eq:GaussChannel}).

\item[(2)]
Alice makes the hard decision.
Then, she obtains the binary data $A_1, \ldots, A_n$.
In this case, $A_i$ and $B_i$ are connected via the binary symmetric channel with crossover probability
$\epsilon_A$, whose mathematical expression will be given later.

\item[(3)]
To send the secret message $M$, using an auxiliary variable $L$, Alice applies the code 
$X^n=(X_1, \ldots, X_n):= \phi_{e,n|S}(M,L) \in \bF_2^n$.
Then, Alice sends ${X'}^n=(X_1', \ldots, X_n')$ to Bob via public channel, where
$X_i':= A_i\oplus X_i$.
Here, an auxiliary variable $L$ is a variable independent of the message $M$,
and is used to realize the secrecy of $M$.

\item[(4)]
Bob decodes $M$ by $\phi_{d,n|S}({X^n}'')$, where  
${X^n}''=(X_1'', \ldots, X_n'')$ and $X_i'':=X_i'\oplus B_i$.
\end{description}

To realize the public channel from Alice to Bob in the second phase,
they employ the RF channel \eqref{eq:SysModel} with 
an error correcting code $(\hat{\phi}_{e,\hat{n}},\hat{\phi}_{d,\hat{n}})$ 
different from the error correction $\phi_{e,n|S}$
so that the decoding error probability of the code $(\hat{\phi}_{e,\hat{n}},\hat{\phi}_{d,\hat{n}})$ is close to zero (i.e., the bit error rate is e.g. below $10^{-6}$).
Here, if the coding rate of the code $(\hat{\phi}_{e,\hat{n}},\hat{\phi}_{d,\hat{n}})$ is $\hat{R}$,
they use the RF channel \eqref{eq:SysModel} $\hat{n}=n/\hat{R}$ times in Step (3) physically.
That is, in Step (3), Alice sends $\hat{X}^{\hat{n}}=\hat{\phi}_{e,\hat{n}} ({X'}^n) $
to Bob via the RF channel \eqref{eq:SysModel} of coefficient $\eta_B$.
Also, in Step (4), to get ${X'}^n$, Bob applies the decoder $\hat{\phi}_{d,\hat{n}}$
to the received $\hat{n}$ symbols via the RF channel \eqref{eq:SysModel} of coefficient $\eta_B$.
Indeed, when the bit error rate of the public channel (i.e., the bit error rate 
of the code $(\hat{\phi}_{e,\hat{n}},\hat{\phi}_{d,\hat{n}})$)
is e.g. below $10^{-6}$,
it can be negligible in comparison with the bit error rate between $A$ and $B$. 
Hence, we can consider that
the bit error rate of the channel from $X_i$ to $X_i''$
given in Steps (3) and (4)
almost equals that between $A$ and $B$, in practice.

The above description has no authentication.
However, it is possible by attaching using universal$_2$ hash function \cite{Wegman1981,Krawczyk},
which prevents active man-in-the-middle attack \cite{Conti2016},
while it is difficult to avoid active man-in-the-middle attacks without authentication \cite{Eberz}.
The detail is discussed in \cite{Fung} and the arXiv version of \cite{Correlation}.

\subsection{Secrecy capacity when Eve uses hard decision}
When Eve has limited memory, it is natural that Eve uses hard detection decoding
when receiving $Z$
so that Eve obtains the binary variable $E$. 

Indeed, the first phase can be regarded as a preparation step for the secure communication.
In order to prevent Eve to make soft decision, Alice and Bob can consider the following strategy.
Before starting the second phase, Alice and Bob continue the first phase so that the length of their obtained random numbers 
$A_1, \ldots, A_n$ and $B_1, \ldots, B_n$ is close to the limitation of their memory.
In this case, the length of their obtained random numbers is across several coding blocks.
Since satellite has a limitation of size of memory due to the limitation of physical space.
it is natural that the size of Eve's memory is similar to that of Alice and Bob.
In this case, it is difficult for Eve to keep the all the outcomes of soft decision, i.e.,
Eve needs to choose hard decision in this case.
For example, the preceding paper \cite{Endo2015}, which is oriented to an application side,
analyzed the security for the Poisson wiretap channel when Eve has limited memory, i.e.,
Eve uses hard detection decoding.

Using two independent Bernoulli random variables $X_1$ and $X_2$ on $\bF_2$,
we have 
\begin{align}
\label{eq:BSC1}
A=B\oplus X_1, \quad
E=B\oplus X_2.
\end{align}
The crossover probability between $A$ and $B$ is $\epsilon_A= 0.5 \mbox{erfc}(\sqrt{\eta_A/2})$ 
and the crossover probability between $E$ and $B$ is $\epsilon_E= 0.5 \mbox{erfc}(\gamma_A \sqrt{\eta_A/2})$,
where $\mbox{erfc}(t)$ is defined in \eqref{LBH}.
Hence, 
the problem is reduced to the case with BSC channels, which was discussed by Maurer \cite{Maurer1993}.
Hence, 
the capacity $C_{\hard}^{\TW}(\gamma_A,\eta_A)$ 
when Eve uses hard detection is calculated to
\begin{align}
&C_{\hard}^{\TW}(\gamma_A,\eta_A) =
I(A;B|E)=
I(A;B)-I(A;E)\nonumber \\
=&
H(B)-H(B|A)-(H(E)-H(E|A)) \nonumber \\
=&H(E|A)-H(B|A)
= h(\epsilon_E +\epsilon_A -2\epsilon_E \epsilon_A)-h(\epsilon_A) 
\end{align}
because $H(B)=H(E)=h(\frac{1}{2})$ and 
the probability of $E\neq A$ is 
$\epsilon_E(1-\epsilon_A) +(1-\epsilon_E)\epsilon_A
=\epsilon_E +\epsilon_A -2\epsilon_E \epsilon_A $.
When $ \gamma_A=\gamma_B$ and $\eta_A=\eta_B$, i.e., 
$\epsilon_B^*=\epsilon_A$
and
$\epsilon_E^*=\epsilon_E$, we have
\begin{align}
C_{\hard}^{\TW}(\gamma_A,\eta_A)
\ge C_{\hard}^{\OW}(\gamma_B,\eta_B)
\end{align}
because $
h(\epsilon_E +\epsilon_A -2\epsilon_E \epsilon_A)
=h(\epsilon_E +\epsilon_A (1-2\epsilon_E))
\ge h(\epsilon_E)=h(\epsilon_E^*)$.

\subsection{Secrecy capacity when Eve uses soft decision}
When Eve has sufficient size of memory
and her computation power is unlimited,
she can employ soft decision decoding.
That is, to consider Eve's best strategy, we need to address the case when Eve
uses the variables $Z$ and $X'$.
To analyze this case, we use the Markov chain $A-B-Z$.
We focus on the wiretap channel composed of the main channel $W_B:= P_{B,X'|X} $ and the eavesdropper channel $W_E:= P_{Z,X'|X} $.
Since $A=X'\oplus X$, we have $P_{B,X'|X} (b,x'|x)= P_{B|A}(b|x'\oplus x)P_A(x'\oplus x)$.
Hence, 
the Markov chain $A-B-Z$ condition guarantees that
\begin{align}
& P_{Z,X'|X}(z,x'|x)\nonumber \\
=&
\sum_{b}P_{Z|B}(z|b) P_{B|A}(b|x'\oplus x)P_A(x'\oplus x) \nonumber\\
=&
\sum_{b}P_{Z|B}(z|b) P_{B,X'|X}(b,x'|x).
\end{align}
Thus, the channel $W_E$ is a degraded channel of the channel $W_B$. 
Further, 
the channels $W_B$ and $W_E$
are symmetric, the wiretap capacity is attained when $P_X$ is the binary uniform distribution,
and the wiretap capacity $C_{\soft}^{\TW}(\gamma_A,\eta_A)$ is calculated to 
\begin{align}
& C_{\soft}^{\TW}(\gamma_A,\eta_A)=I(X;X'B)-I(X;X'Z) \nonumber \\
=& I(X;X')+I(X;B|X')-(I(X;X')+I(X;Z|X'))\nonumber \\
=& I(A\oplus X';B|X')-I(A\oplus X';Z|X')\nonumber \\
=& I(A;B|X')-I(A;Z|X')
\stackrel{(a)}{=} I(A;B)-I(A;Z)\nonumber \\
=& I(A;BZ)-I(A;Z)= I(A;B|Z),
\end{align}
where $(a)$ follows from the independence of $X'$ from $A,B,Z$,
which can be shown by the uniformity of the conditional distribution $P_{X'|A}$.
Therefore, the wiretap capacity $C_{\soft}^{\TW}$ is always positive regardless of $\gamma_A$, 
regardless the condition $\gamma_A <1$ does not hold. 
The wiretap capacity expresses the limit of the secure transmission rate 
when we use a proper coding under the condition that
the mutual information between the message and Eve's information goes to zero.

Further, 
the probability $P_Z$ and 
the conditional probability $P_{B|Z}$ are calculated as
\begin{align}
P_{Z}(z)=&
\sum_{b=0}^1 P_{Z,B}(z,b)=
\sum_{b=0}^1 P_{Z|B}(z|b)P_B(b) \nonumber\\
=&
\frac{1}{2\sqrt{2 \pi }}
(e^{-\frac{(z+\gamma_A\sqrt{\eta_A})^2}{2}}
+e^{-\frac{(z-\gamma_A\sqrt{\eta_A})^2}{2}}),\\
P_{B|Z}(b|z)=& 
\frac{P_{B,Z}(b,z)}{P_Z(z)}
= \frac{P_{Z|B}(z|b) P_B(b)}{ P_Z(z)}
= \frac{P_{Z|B}(z|b)}{2 P_Z(z)}\nonumber \\
= &
\frac{e^{-\frac{(z-(-1)^b\gamma_A \sqrt{\eta_A})^2}{2}}
}
{
e^{-\frac{(z+\gamma_A\sqrt{\eta_A})^2}{2}}
+e^{-\frac{(z-\gamma_A\sqrt{\eta_A})^2}{2}}
}.
\end{align}
Hence, 
due to the Markov chain $Z-B-A$,
we can calculate the conditional mutual information; 
\begin{align*}
 I(A;B|Z=z) 
 =
 h\Big(P_{B|Z}(0|z) \epsilon_A+ P_{B|Z}(1|z)(1-\epsilon_A)\Big)-h(\epsilon_A).
\end{align*}
Notice that $ I(A;B|Z=z) =0$ if and only if 
$P_{B|Z}(0|z)$ is $0$ or $1$.
Hence, 
the capacity $C_{\soft}^{\TW}(\gamma_A,\eta_A)$ with Eve's soft decision 
is calculated as a function of $\eta_A,\gamma$ by 
\begin{align}
&C_{\soft}^{\TW}(\gamma_A,\eta_A)= I(A;B|Z) \nonumber \\
 =&
\int_{-\infty}^{\infty}
P_Z(z)dz
\Big( h\Big(  P_{B|Z}(0|z) \epsilon_A+ P_{B|Z}(1|z)(1-\epsilon_A)\Big)
-h(\epsilon_A)
\Big).\label{LLF}
\end{align}
Thus, unless $P_{B|Z}(0|z)$ is $0$ or $1$ for all $z$, 
\eqref{LLF} is strictly positive.
That is, when $ \gamma_A$ is a finite value, 
$P_{B|Z}(0|z)$ is an intermediate value between $0$ and $1$ for all $z$.
Hence, the capacity $C_{\soft}^{\TW}(\gamma_A,\eta_A)$ is strictly positive.

\begin{figure}[tbh]
\centering
\includegraphics[scale=0.3]{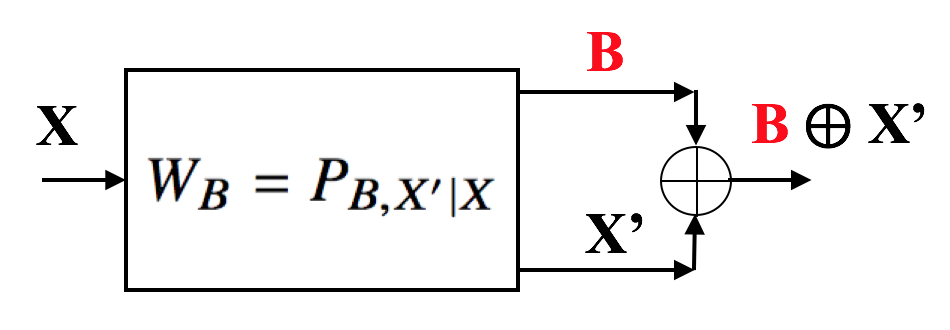}
\protect\caption{Bob's computation channel model for the two-way protocol. Independently of whether Eve uses hard or soft detection, two-way secrecy capacity can be attained with this computation model.}
\label{fig:ComputationChannel}
\end{figure}

Even in this scenario, 
when $X$ is also subject to the uniform distribution,
$B\oplus X'$ is Bob's sufficient statistics with respect to $X$.
Hence, we have
$I(X;X'\oplus B)-I(X;X'Z)
=I(X;X'B)-I(X;X'Z) 
=C_{\soft}^{\TW}(\gamma_A,\eta_A)$.
Therefore, even when Bob uses only $B\oplus X' $ for his decoding 
while Eve uses the two variables $Z$ and $X'$,
the capacity $C_{\soft}^{\TW}(\gamma_A,\eta_A)$ can be attained.
Consequently, the channel $W_B$ can be modeled as a computation channel as illustrated in Fig. \ref{fig:ComputationChannel}.

\begin{figure}[tbh]
\centering
\includegraphics[scale=0.17]{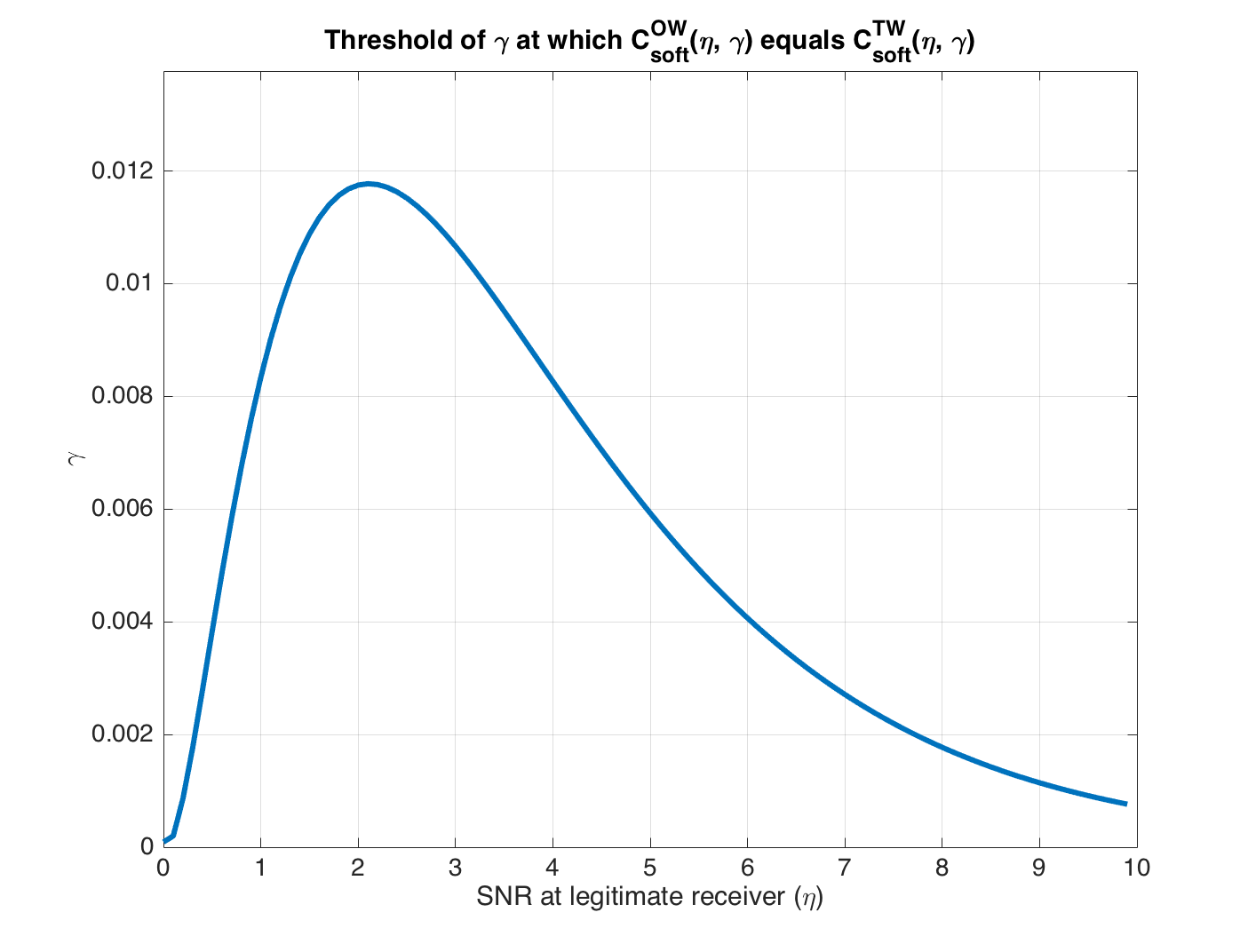}
\protect\caption{Threshold values of $\gamma$ at which $C_{\soft}^{\OW}(\gamma,\eta)$ equals $C_{\soft}^{\TW}(\gamma,\eta)$.}
\label{fig:umbralesX}
\end{figure}

Now, we compare 
$C_{\soft}^{\TW}(\gamma_A,\eta_A)$
and $C_{\soft}^{\OW}(\gamma_B,\eta_B)$
when $ \gamma_A=\gamma_B=\gamma$ and $\eta_A=\eta_B=\eta$.
For this comparison, we fix $\eta$ and change $\gamma$.
Due to the above discussion, when $\gamma>1$, 
$C_{\soft}^{\TW}(\gamma,\eta)$ is larger than $C_{\soft}^{\OW}(\gamma,\eta)$.
In contrast, under the limit $\gamma\to 0$, we have
\begin{align}
& \lim_{\gamma\to 0}C_{\soft}^{\TW}(\gamma,\eta)
=\! \log 2 - h(\epsilon_A) 
=\! I(B_{\TW};A_{\TW})
<  I (B_{\TW};Y_{\TW})
\nonumber\\
= &I (A_{\OW};Y_{\OW}) =I (X_{\OW};Y_{\OW}) 
=\lim_{\gamma\to 0}C_{\soft}^{\OW}(\gamma,\eta),
\label{CapPeor}
\end{align}
where the subscripts $\TW$ and $\OW$ of  the random variables 
express the protocol to be considered.
This opposite inequality is caused by the hard decision on Alice's received signal $Y$ in the two way protocol.
Therefore, when $\gamma$ is smaller than a certain threshold, $C_{\soft}^{\TW}(\gamma,\eta)$ is smaller than $C_{\soft}^{\OW}(\gamma,\eta)$.
That is, the one-way protocol may have greater capacity than the two-way protocol for some threshold of $\gamma$. 
We have computed the value of such threshold as a function of the SNR, which is shown in Fig. \ref{fig:umbralesX}.


\section{Optimization}
Here, to extract a higher communication speed,
we consider how to optimize the channel parameters
in the RF channel \eqref{eq:GaussChannel}.
When the power of transmitting antenna of Bob increases,
the coefficient ${\eta_A}$ increases and
the ratio between the coefficients of signal in Alice's and Eve's sides is not changed.
For simple analysis, we first assume that 
Alice and Bob can know the value of $\eta_A$ by using test transmission,
and control it by changing the power of transmitting antenna of Bob,
where other components (e.g., the receiving antenna gains of Alice and Eve, the directions of antennas etc)
are fixed.
In practice, it is not so easy to know the value of the ratio $\gamma_A$
because it depends on Eve's position.
However, if we know the type of Eve's orbit, we know the range ${\cal G}$ of possible values of $\gamma_A$.
In this case, we consider the worst case for Alice and Bob, i.e., 
$\gamma_{A,\max}:=\max_{\gamma_A \in {\cal G}} \gamma_A$.
In fact, when we have two possible values $\gamma_{A,1}>\gamma_{A,2}$ for the ratio,
the channel to Eve with $\gamma_{A,2}$ is a degraded channel of the channel to Eve with $\gamma_{A,1}$.
Hence \footnote{Note that in a traditional communication scenario, the transmission power is designed so that the link budget can provide a required link quality (e.g. in terms of target bit error rate) is met. However, here we are designing a secure communication scenario and therefore the link budget is constrained to meet the security requirements. Such security requirement means here that the link budget is designed to provide the maximum secrecy capacity.}, a secure code for the channel to Eve with $\gamma_{A,1}$
is also secure for the channel to Eve with $\gamma_{A,2}$.
Therefore, it is sufficient to prepare a code with the largest value 
$\gamma_{A,\max}$.
We optimize 
$C_{\hard}^{\TW}(\gamma_{A,\max},\eta_A)$ and 
$C_{\soft}^{\TW}(\gamma_{A,\max},\eta_A)$ by changing $\eta_A$.
Here, the parameter $\eta_A=P/N_A$ can be changed by changing the power $P$.
The optimum secret capacities are given as
\begin{align}
C_{\soft}^{\TW}(\gamma_{A,\max})&:=\max_{\eta_A}C_{\soft}^{\TW}(\gamma_{A,\max},\eta_A) \\
C_{\hard}^{\TW}(\gamma_{A,\max})&:=\max_{\eta_A}C_{\hard}^{\TW}(\gamma_{A,\max},\eta_A).
\end{align}
Hence, we need to find suitable value for $\eta_A$ dependently of 
$\gamma_{A,\max}$.
$\eta_{\soft}^{\TW}(\gamma_{A,\max}):= \argmax_{\eta_A}C_{\soft}^{\TW}(\gamma_{A,\max},\eta_A)$
and
$\eta_{\hard}^{\TW}(\gamma_{A,\max}):= \argmax_{\eta_A}C_{\hard}^{\TW}(\gamma_{A,\max},\eta_A)$ 
are the optimal intensities of $\eta_A$.

In fact, we can apply a similar optimization to the one way case.
In this case, we consider the following optimum secret capacities;
\begin{align}
C_{\soft}^{\OW}(\gamma_{B,\max})& :=\max_{\eta_B}C_{\soft}^{\OW}(\gamma_{B,\max},\eta_B)\\
C_{\hard}^{\OW}(\gamma_{B,\max})&:=\max_{\eta_B}C_{\hard}^{\OW}(\gamma_{B,\max},\eta_B),
\end{align}
where $\gamma_{B,\max}$ is the maximum value among possible values of 
$\gamma_{B}$.
$\eta_{\soft}^{\OW}(\gamma_{B,\max}):= \argmax_{\eta_B}C_{\soft}^{\OW}(\gamma_{B,\max},\eta_B)$
and
$\eta_{\hard}^{\OW}(\gamma_{B,\max}):= \argmax_{\eta_B}C_{\hard}^{\OW}(\gamma_{B,\max},\eta_B)$ 
are the optimal intensities of $\eta_B$.

Now we show numerical calculations to compare the capacities of the one-way protocol and the two-way protocol. For easy visualization, we calculate numerical values assuming $\eta_B=\eta_A$ and $\gamma_A=\gamma_B$.

\begin{figure}[tbh]
\centering
\includegraphics[scale=0.2]{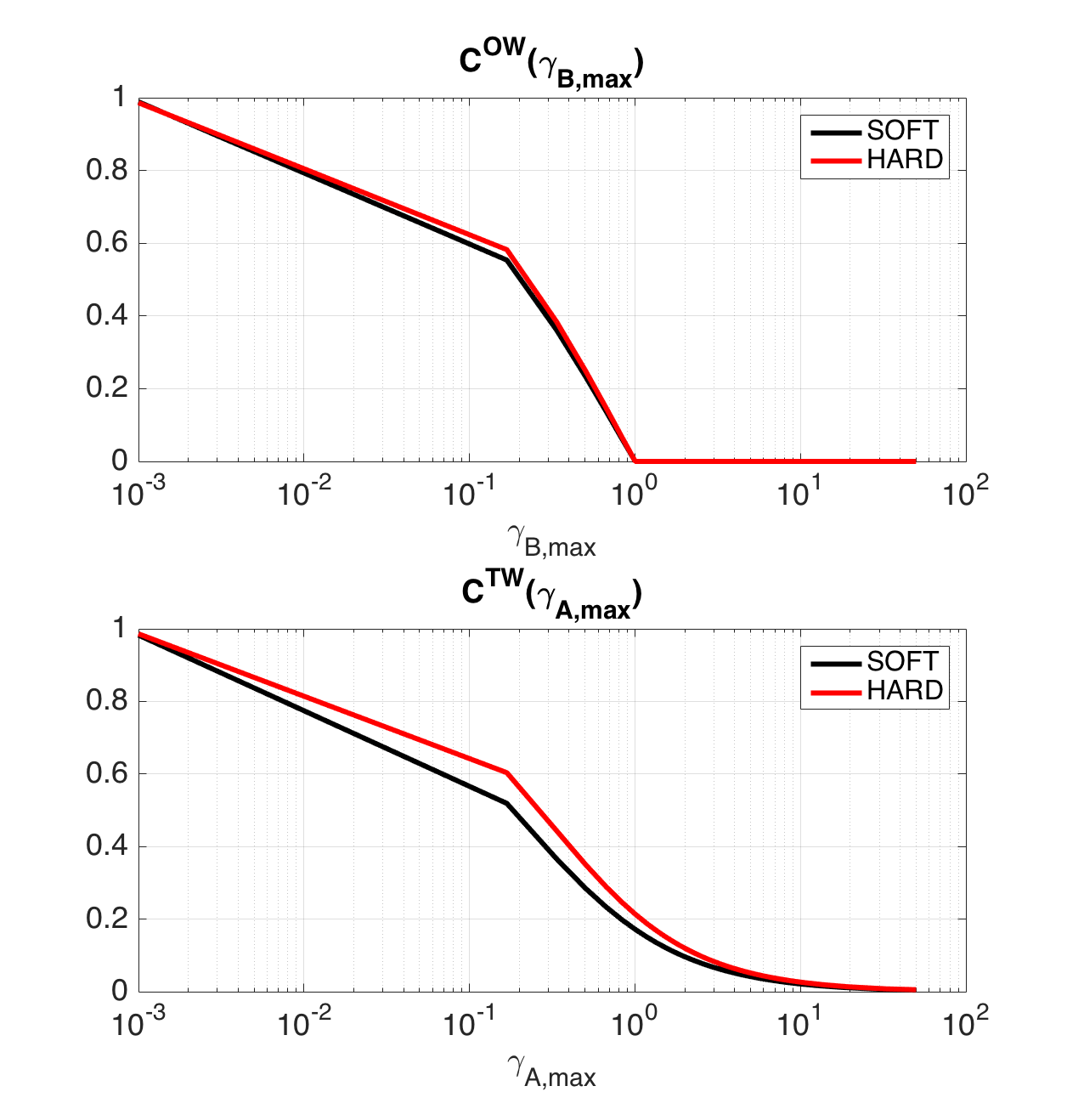}
\caption{Comparison among the optimum secret capacities
$C_{\soft}^{\TW}(\gamma_{A,\max})$,
$C_{\hard}^{\TW}(\gamma_{A,\max})$,
$C_{\soft}^{\OW}(\gamma_{B,\max})$, and
$C_{\hard}^{\OW}(\gamma_{B,\max})$.
The vertical axis expresses these optimal capacities.
The horizontal axis expresses 
$\gamma_{A,\max}$ and $\gamma_{B,\max}$ with log scale, which runs from $10^{-3}$ to $10^2$.}
\label{fig:1_Comparison}
\end{figure}

\begin{figure}[tbh]
\centering
\includegraphics[scale=0.17]{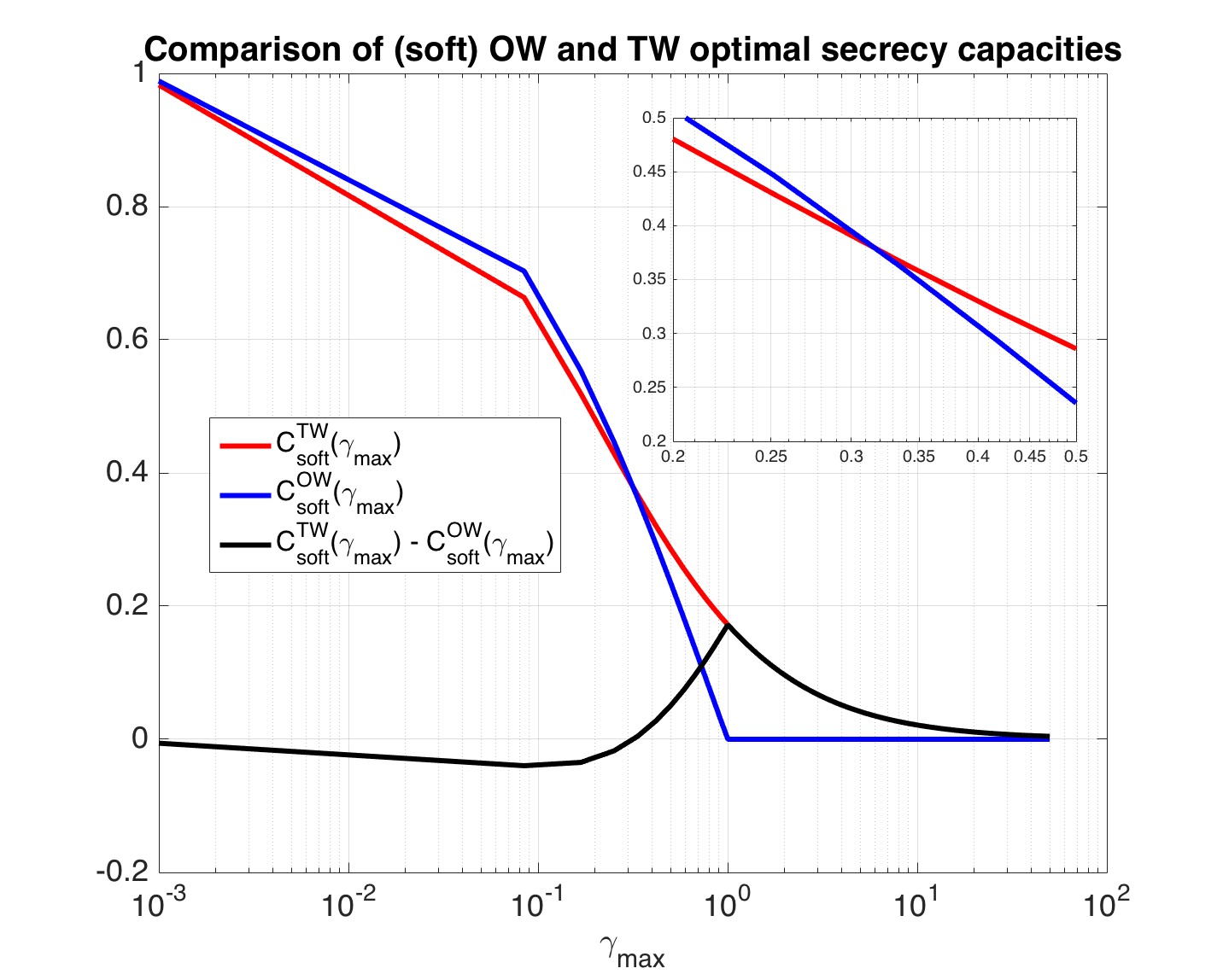}
\protect\caption{Comparison of OW and TW optimal secrecy capacities assuming $\gamma_{B,\max}=\gamma_{A,\max}=\gamma_{\max}$.}
\label{fig:umbrales}
\end{figure}

Fig. \ref{fig:1_Comparison} shows the comparison 
among the optimum secret capacities
$C_{\soft}^{\TW}(\gamma_{A,\max})$,
$C_{\hard}^{\TW}(\gamma_{A,\max})$,
$C_{\soft}^{\OW}(\gamma_{B,\max})$, and
$C_{\hard}^{\OW}(\gamma_{B,\max})$.
We can observe that while for the one-way protocol the capacity is zero whenever the eavesdropper has higher SNR than Bob, in the two-way protocol the capacity is always positive and greater than zero. We have computed the difference between optimal secrecy capacities $C_{\soft}^{\OW}(\gamma_{B,\max})$ and $C_{\soft}^{\TW}(\gamma_{A,\max})$ for $\gamma_{B,\max}=\gamma_{A,\max}=\gamma_{\max}$
as Fig. \ref{fig:umbrales}. 
We observe that as obtained theoretically in \eqref{CapPeor}, the OW secrecy capacity is slightly bigger than the TW secrecy capacity when the channel to Bob is advantageous over that to Eve.
However, also in agreement with the theoretical derivations, the TW secrecy capacity starts to be bigger than 
the OW 
when the channel to Bob is not so advantageous over that to Eve.
In  Fig. \ref{fig:umbrales}, we observe in the zoom plot that this occurs at $\gamma_{max}=0.3185$.

\begin{figure}[tbh]
\centering
\includegraphics[scale=0.22]{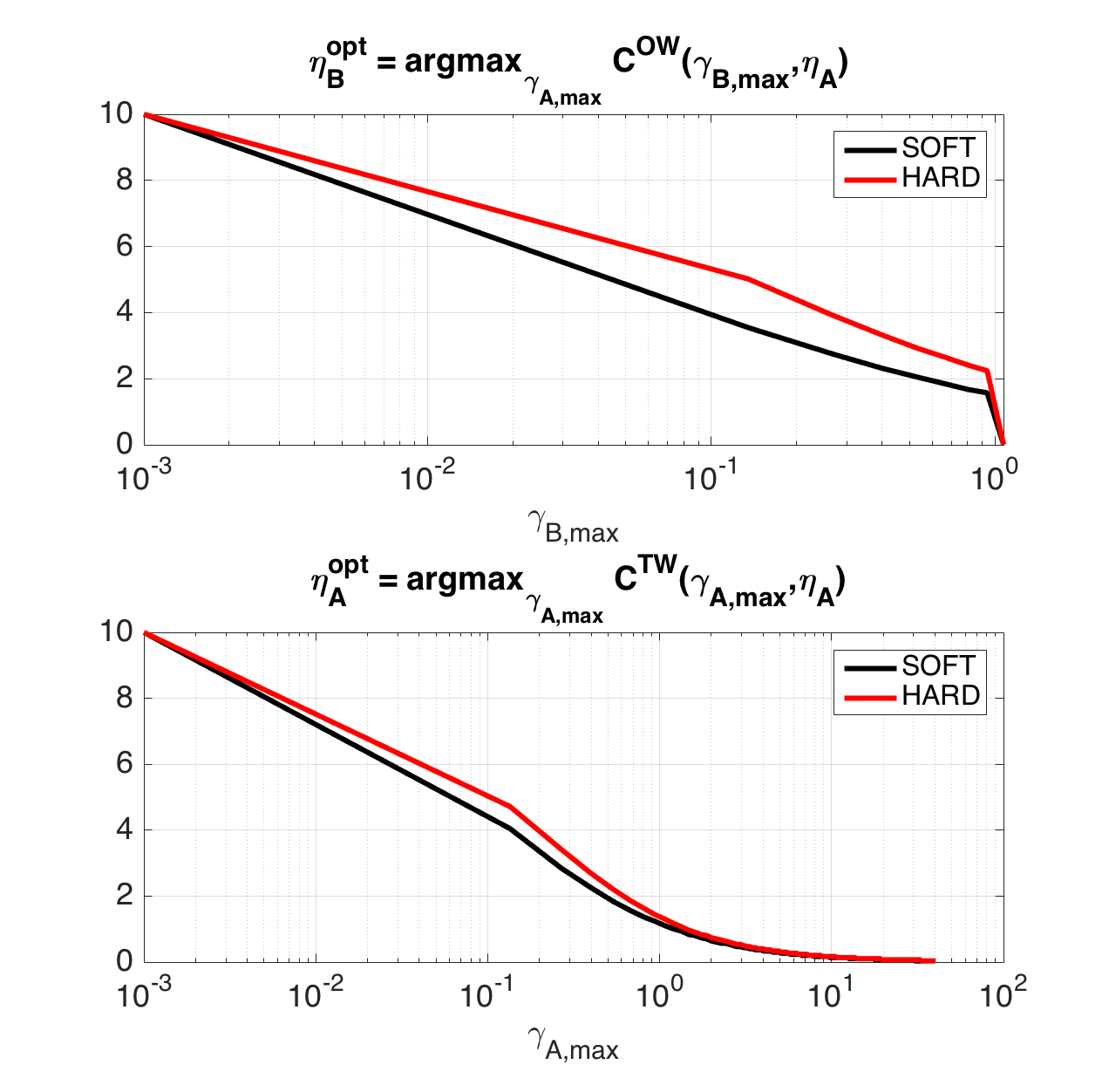}
\protect\caption{
Optimal intensities
$\eta_{\soft}^{\TW}(\gamma_{A,\max})$,
$\eta_{\hard}^{\TW}(\gamma_{A,\max})$,
$\eta_{\soft}^{\OW}(\gamma_{B,\max})$, and
$\eta_{\hard}^{\OW}(\gamma_{B,\max})$.
The vertical axis expresses these optimal intensities with log scale.
The horizontal axis expresses 
$\gamma_{A,\max}$ and $\gamma_{B,\max}$ with log scale,
where
$\gamma_{A,\max}$ runs from $10^{-3}$ to $10^2$, but 
$\gamma_{B,\max}$ runs from $10^{-3}$ to $1$.
}
\label{fig:2_Comparison}
\end{figure}

Fig. \ref{fig:2_Comparison} shows 
the optimal intensities
$\eta_{\soft}^{\TW}(\gamma_{A,\max})$,
$\eta_{\hard}^{\TW}(\gamma_{A,\max})$,
$\eta_{\soft}^{\OW}(\gamma_{B,\max})$, and
$\eta_{\hard}^{\OW}(\gamma_{B,\max})$.
These values are the optimal choices for the intensity $\eta_A$ or $\eta_B$ in the respective cases.

\section{Application to real satellite communication}
\begin{figure}[tbh]
\centering
\includegraphics[scale=0.4]{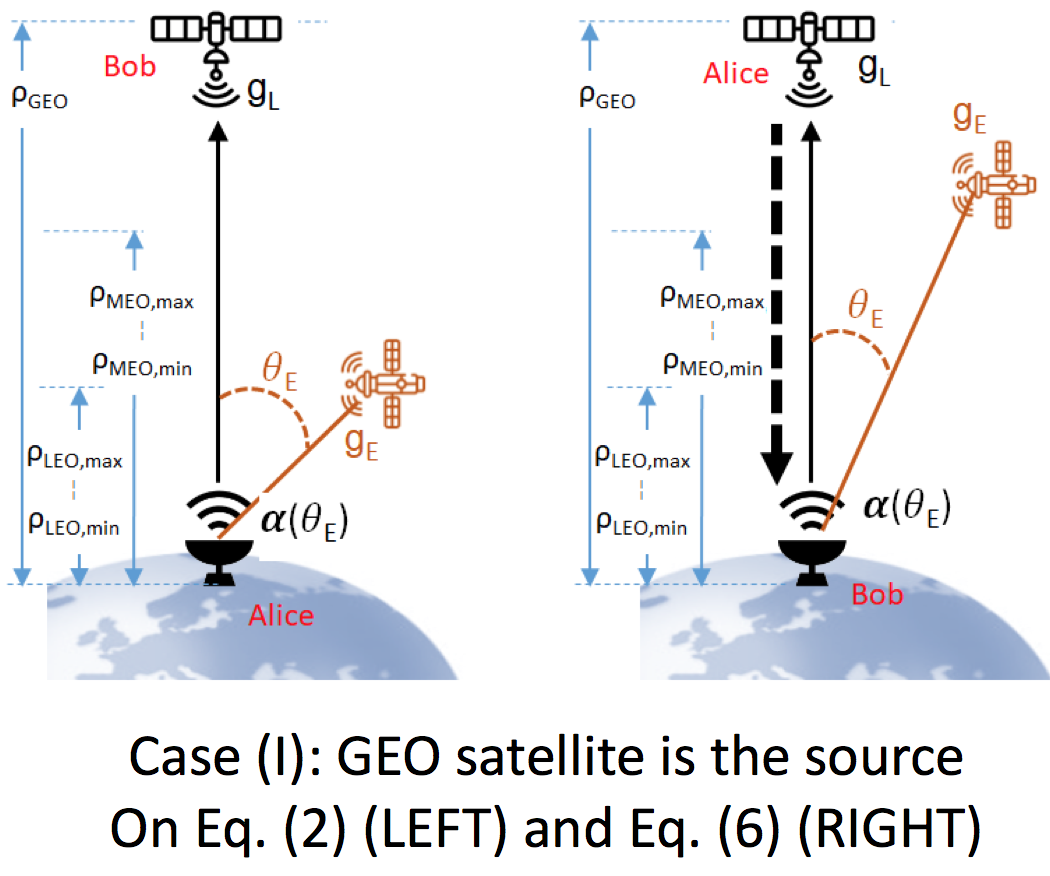}
\includegraphics[scale=0.4]{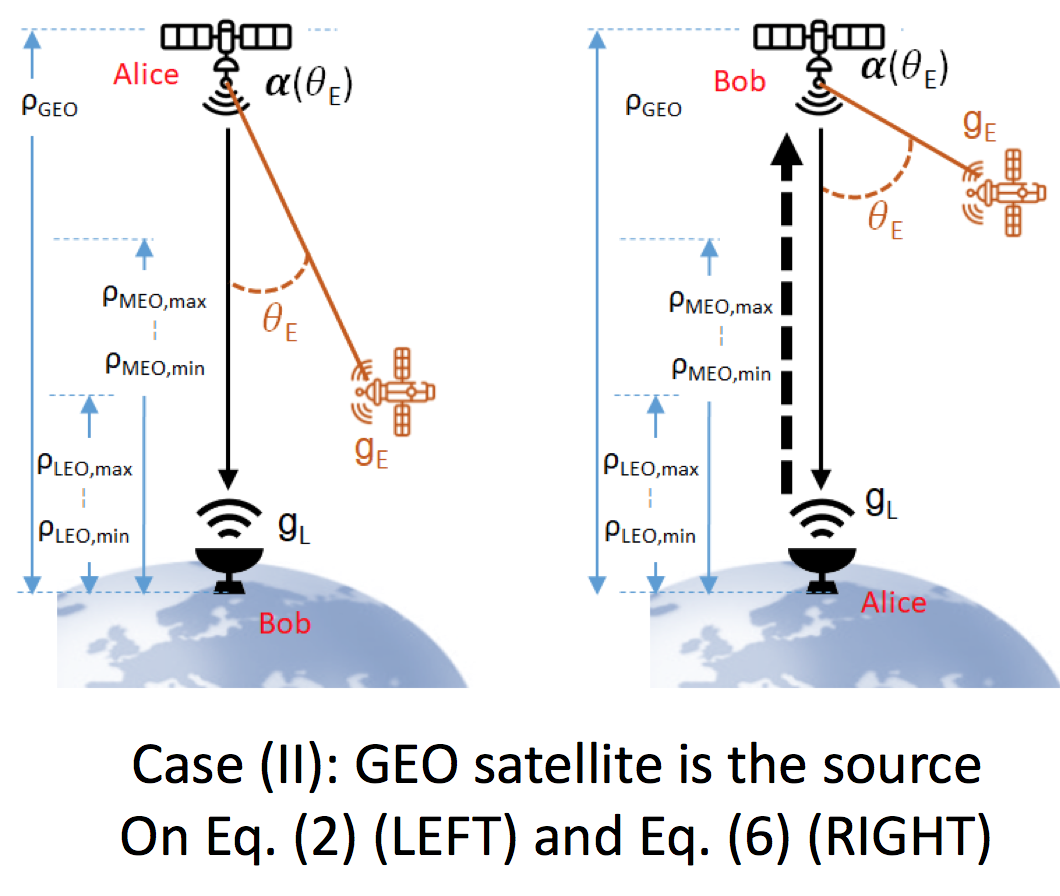}

\protect\caption{Geometry considered to illustrate the real satellite communication secrecy analysis. 
The two figures on the top describe Case (I), in which the Earth station is the source 
in the noisy Gaussian channels \eqref{eq:SysModel} and \eqref{eq:GaussChannel}.
The two figures on the bottom describe Case (II), in which the satellite is the source 
in the noisy Gaussian channels \eqref{eq:SysModel} and \eqref{eq:GaussChannel}.
The geometries for OW (left) and TW (right) cases illustrate LEO and MEO orbit heights. 
The dashed line expresses the noiseless public channel from Alice to Bob in the TW case.
The eavesdropper can be at any LEO or MEO orbit while the legitimate transmitter and receiver are either the earth station of the GEO satellite. }
\label{fig:geometry}
\end{figure}

Now, we apply our analysis to the following two types of real satellite communication scenarios. 
\begin{description}
\item[(I)]
The transmitter is the earth station and 
the legitimate receiver is the GEO satellite in the noisy Gaussian channels 
\eqref{eq:SysModel} and \eqref{eq:GaussChannel}.
That is, Alice is the earth station and Bob is the GEO satellite in the OW, 
and 
Bob is the earth station and Alice is the GEO satellite in the TW. 
Notice that
the noiseless public channel from the GEO satellite to the earth station
is also required by using a proper combination of wireless communication and 
outer error correcting code in the TW.
The noisy Gaussian channels \eqref{eq:SysModel} and \eqref{eq:GaussChannel}
of these scenarios 
are explained in the two figures on the top in Fig \ref{fig:geometry}, which describe the case when the Earth station is the information data communication source in the noisy Gaussian channel. 
In this case, the eavesdropper, Eve is a low Earth orbit (LEO) satellite
or a medium Earth orbit (MEO) satellite.

\item[(II)]
The transmitter is the GEO satellite and 
the legitimate receiver is the earth station in the noisy Gaussian channels \eqref{eq:SysModel} and \eqref{eq:GaussChannel}.
That is, Alice is the GEO satellite and Bob is the earth station in the OW, 
and 
Bob is the GEO satellite and Alice is the earth station in the TW. 
The noisy Gaussian channels \eqref{eq:SysModel} and \eqref{eq:GaussChannel} of these scenarios 
are explained in the two figures on the down in Fig \ref{fig:geometry}, 
which describe the case when the GEO satellite is the information data communication source in the noisy Gaussian channel. 
In this case, Eve is 
an LEO satellite, an MEO satellite, or a GEO satellite.
\end{description}

 
Let $\alpha(\theta)$ be the normalised transmitter's antenna radiation's pattern of the earth station
in response to the angle $\theta$ from the boresight axis directed to 
the GEO satellite to account for spatial attenuation. 
The function $\alpha(\theta)$ can be considered exactly in case the 
(normalized) antenna pattern is known, 
or otherwise it can be considered in terms of the allowed emission of radiation according to space regulations. A typical analytical expression for $\alpha(\theta)$ is
\begin{equation}\label{H31}
\alpha(\theta) := \frac{J_1(k\sin(\theta))}{2k\sin(\theta)} 
+ 36\frac{J_3(k\sin(\theta))}{(k\sin(\theta))^3}
\end{equation}
where $k = 2.0712/\sin(\theta^{3dB})$, with $\theta^{3dB}$ 
being the one-sided half-power angular beamwidth and $J_1$ and $J_3$ are 
the Bessel functions of the first kind, of order one and three respectively. 
Our interest is the parameter $\alpha(\theta_E)$ in the specific angle $\theta_E$ between Bob's and Eve's directions. 
$g_{L}$ and $g_{E}$ are legitimate and eavesdropper's receiver's antenna gains towards Earth station. 
Now we introduce a model for the coefficient that gives Eve's signal strength relative to Bob's signal in \eqref{eq:SysModel} and \eqref{eq:GaussChannel} (see \cite{PhySecSat}\cite{arXiv}\cite{RealisticChannel}). We first introduce the parameter $\mu$ to account for the relative antenna gain, i.e.,
$\mu:= \sqrt{g_{L}/g_{E}}$.
We also define $\beta\left(r,\rho_{E}\right)$ to account for relative propagation losses between Bob and Eve as
\begin{align}\label{H33}
\beta^2\left(r,\rho_{E}\right)=\frac{\rho_{L}^{2}}{\rho_{E}^{r}}.
\end{align}
The exponent $r$ accounts for the power attenuation decay that affects eavesdropper's propagation channel. Different values of the exponent model
correspond to different assumptions about eavesdropper. 
Specifically, the eavesdropper can be modeled as a terrestrial, aerial or satellite station. 
For example, while for the satellite case $r=2$, in case of aerial eavesdropper, 
a good assumption is to consider a large scale two-ray ground multipath model,with $r>2$.
Then, we discuss 
the parameter $\gamma_B$ in the OW \eqref{eq:SysModel} and 
the parameter $\gamma_A$ in the TW \eqref{eq:GaussChannel}
because 
the channel \eqref{eq:SysModel} in the OW case
is the same as
the channel \eqref{eq:GaussChannel} in the TW case
in each scenario (I) or (II).
These parameters are given as
\begin{align}
\gamma(\theta_E, \rho_E, r,\mu,\gamma_n): = 
\frac{\alpha\left(\theta_{E}\right)
\mu\beta\left(r,\rho_{E}\right)}{\sqrt{\gamma_n}},
\label{grrt}
\end{align}
where, $\gamma_n$ is the ratio between the powers of the noises in 
legitimate receiver's and eavesdropper's detectors.
In doing a secrecy analysis, we assume in which orbit Eve is, 
but we don't make any assumption on which angle she is (since she is orbiting).
In this case, to guarantee the security, 
we need to consider the worst case.
For this aim, we consider the possible range ${\cal R}$ of the value $(\theta_E, \rho_E)$.
Then, the maximums of $\gamma_B$  and $\gamma_A$ are calculated to 
\begin{align}
\gamma_{B,\max}=\gamma_{A,\max}=
\max_{(\theta_E, \rho_E) \in {\cal R}}
\gamma(\theta_E, \rho_E, r,\mu,\gamma_n).
\end{align}

Now, we assume representative values for Eve's possible orbits according to basic orbital mechanics \cite{OrbitMechanics} and usual low or medium orbit terminology. In Case (I), 
when Eve is MEO (LEO), we assume that the height of Eve's orbit runs from 
$\rho_{\MEO\min}=5000$ to $\rho_{\MEO\max}=20000$ km 
($\rho_{\LEO\min}=150$ to $\rho_{\LEO\max}=2000$ km).
Also, we assume that the height of our GEO orbit is $\rho_{\GEO}= 36000$ km.
Since the maximum of $\alpha\left(\theta_{E}\right)$ is realized by $\theta_{E}=0^{\circ}$,
when Eve is MEO (LEO), 
we have
$\gamma(0^{\circ}, \rho_E, r,\mu,\gamma_n)
= \frac{\mu \rho_{\GEO}}{ \rho_{\MEO\min}^{r/2}\sqrt{\gamma_n}}$
($= \frac{\mu \rho_{\GEO}}{ \rho_{\LEO\min}^{r/2}\sqrt{\gamma_n}}$).
For illustration, we now assume Eve equally powerful than the legitimate receiver, i.e. $\mu=1$ and $\gamma_n=1$. Also, we have $r=2$ for both LEO and MEO.
Hence, the maximum $\gamma_{B,\max}=\gamma_{A,\max}$
is given as 
$\gamma_{{\rm (I)}\MEO}:=
\frac{ \rho_{\GEO}}{ \rho_{\MEO\min} }= 36000 / 5000 = 7.2$
for the case when Eve is MEO,
and it is given as
$\gamma_{{\rm (I)}\LEO}:=
\frac{ \rho_{\GEO}}{ \rho_{\LEO\min} }= 36000/150 = 240$
for the case when Eve is LEO.
Then, we have the capacities of the worst case as Table \ref{T1}.
While these capacities are small in comparison with the conventional communication,
we see that the secure communication is possible in these scenarios only in the TW case.

Applying same reasoning in Case (II), when Eve is MEO,
the minimum of $\rho_E$ is $\rho_{\GEO}-\rho_{\MEO\max}$ at $\theta_E=0^{\circ}$
and the maximum of $\alpha\left(\theta_{E}\right)$ is realized by $\theta_{E}=0^{\circ}$.
The same observation holds when Eve is LEO or GEO.
In this case, it seems
reasonable to assume the legitimate receiver having a more
powerful antenna gain and less detector noise than the eavesdropper.
Hence, we can again assume $\mu=1$ and $\gamma_n=1$.
Again, we also have $r=2$.
Therefore, when Eve is MEO,
the maximum $\gamma_{B,\max}=\gamma_{A,\max}$
is calculated to be
$\gamma_{{\rm (II)}\MEO}:=
\frac{\rho_{\GEO}}{ (\rho_{\GEO}-\rho_{\MEO\max})}=36000/(36000-20000)=9/4$.
When Eve is LEO,
it is calculated to be
$\gamma_{{\rm (II)}\LEO}:=
\frac{ \rho_{\GEO}}{ (\rho_{\GEO}-\rho_{\LEO\max}) }=36000/(36000-2000)=18/17$.
When Eve is GEO,
the minimum distance between the transmitter of the initial transmission and Eve 
is 1km.
Hence, 
it is calculated as
$\gamma_{{\rm (II)}\GEO}:=
\frac{\rho_{\GEO}}{ 1 }=36000$.

\begin{table}[htpb]
  \caption{Summary of optimum capacities in satellite models}
\label{T1}
\begin{center}
  \begin{tabular}{|l|c|c|c|c|}
\hline
$\gamma_{B,\max}=\gamma_{A,\max}$
 & $C_{\soft}^{\TW}$ &  $C_{\hard}^{\TW}$ & $C_{\soft}^{\OW}$ &  $C_{\hard}^{\OW}$ \\
\hline
$\gamma_{{\rm (I)}\MEO}$ & $0.029$ & $0.036$&  0&0 \\
\hline
$\gamma_{{\rm (I)}\LEO}$ & $3.6  \times 10^{-4}$ & $3.9  \times 10^{-4}$ & 0&0 \\
\hline
$\gamma_{{\rm (II)}\GEO}$ & $1.4  \times 10^{-12}$ & $1.4  \times 10^{-12}$ &  0&0 \\
\hline
$\gamma_{{\rm (II)}\MEO}$ & $0.086$ & $0.108$ &  0&0 \\
\hline
$\gamma_{{\rm (II)}\LEO}$ & $0.159$ & $0.198$ &  0&0\\
\hline
  \end{tabular}
\end{center}
\end{table}

\section{Conclusions and further improvements}
We have introduced a two-way protocol for the BPSK modulation
to overcome the limitations of the classic (one-way) wiretap physical layer security protocol. 
While the secrecy capacity of the one-way protocol is negative when Eve's channel is better than Bob's channel (i.e $\gamma_B > 1$), we show that the two-way protocol always provides positive capacity with higher gains even for 
$\gamma_A \ge 1$ and $\gamma_B \ge 1$ when the noises exist and are independent.
We have shown that 
the one-way protocol cannot realize secure communication in realistic scenarios of satellite communication,
while our two-way protocol can realize secure communication in these realistic scenarios.
Notice that this conclusion does not change whenever the maximum of possible values of
$\gamma_A $ and $\gamma_B $ is greater than $1$.

For example, in the scenario (I), we have the transmission rate 3.6$\times 10^{-4}$ with the worst case analysis
in the two-way protocol with Eve's soft decision
while the eavesdropper has an extremely stronger power in the detection process than the legitimate receiver, i.e., the eavesdropper has 240 times power in the receiving signal as the legitimate receiver. 
The conventional one-way method cannot realize secure communication in this case.
This numerical analysis shows that even in this case,
we can realize secure communication with the same physical device
if we accept approximately e.g. one thousandth reduction of the speed of the conventional communication without secrecy (e.g., 1 Gbps is reduced to 1 Mbps).
While this cost seems very large,
the cost is still much smaller than quantum key distribution (QKD) \cite{BB84} due to the following reason.
Since QKD requires expensive devices, it is available only for extremely limited users 
(big governments and/or big military organizations).
However, since the proposed method is based on the conventional satellite system,
even though the transmission speed is very low, 
it is available for ordinary users.
In fact, the user can use the proposed system 
when the size of communication is reduced, e.g., the user uses only e-mail instead of video.
On the other hand, in the scenario (II), 
we assume that the receiving antenna of the earth station has almost same performance 
as that of Eve.
Under such a worst case assumption, we obtain very small transmission speed.
To improve this, the receiving antenna of the earth station needs to be more powerful than that of Eve.
However, to realize this condition,
the earth station needs to prepare an expensive receiving device, which may or may not imply to restrict ordinary users since such higher cost is shared by all service/users sharing the earth station.
In this sense, the scenario (II) may seem less practical for ordinary users.
In any case, to realize secure communication, it is sufficient to share secure keys between two users.
because one-time use of shared secure key realizes secure communication with both directions.
Hence, it is sufficient to establish secure communication only with one direction.
Therefore, the scenario (I) of TW is enough for our purpose.

However, in the scenario (I) of TW, if Eve is an eavesdropping terrestrial node, e.g., a drone near the terrestrial earth station, she has better performance than LEO/MEO satellite.
In this case, the secure transmission rate is worse than 3.6$\times 10^{-4}$
when the angle $\theta_E$ is set to be $0$.
However, the possible minimum angle $\theta_E$ of this case in practice is larger than 
that in the case with an eavesdropping LEO/MEO satellite.
Therefore, it is needed to evaluate the secure transmission rate with 
an eavesdropping terrestrial node and 
the possible minimum angle $\theta_E$.
However, it is not so easy to find the minimum angle $\theta_E$
among practically possible values.
Therefore, this type of analysis is remained as a future study.

As the price to pay, the protocol requires higher delay to establish the secure channel when compared to the one-way. 
However, this cost is much cheaper than 
the previous two-way protocols in the papers \cite{WGH,FJHWY,ZWPT,TY} 
because 
they require many rounds of communication while our protocol requires only two rounds of communication.
On the other hand, the transmission of information can be over a public channel while for randomness sharing, the channel needs to be previously authenticated like \cite{Correlation,Fung}. 
As discussed in \cite{Correlation,Fung}, the required amount of the random numbers shared  between Alice and Bob in advance is the logarithm order of the size of intended secure communication.
Also, this cost is much cheaper than the realization of quantum key distribution. Therefore, considering the cost-benefit performance, we find that our two-way method is useful.
Furthermore, the bias of the variable $B$ may reduce the effectiveness of the protocol and reduce the secrecy capacity gains. 
To improve this problem, we often distill uniform random numbers from the thermal noise.
It is known that it is possible to distill uniform random numbers 
by applying a hash function to a biased random numbers \cite{BBCM,HILL,Hayashi2011,H13}.
To obtain the ultimate secure uniform random number, we may employ quantum random number generator \cite{HG,MYQZ,HZ},
which requires much cheaper cost than 
quantum key distribution because it does not need quantum communication.

Unfortunately, this paper discusses only the asymptotic performance.
Since the implemented communication system has finite-length codes,
we need to evaluate the security of finite-length codes for its practical application \cite{ISIT2013,HTW1,HTW2,YSP,HarXiv}.
Since the finite-length analysis depends on the choice of the security criterion,
we need to be careful of its choice \cite{hay-wire,H13}.
As such a study is beyond the focus on this paper, 
it is considered as a future study. 
Furthermore, it is well known that a good model for the land mobile satellite (LMS) channel model is a Markov model \cite{Fontan}. 
Hence, it is a completely different channel from \eqref{eq:GaussChannel}.
Therefore, follow up studies also include considering different satellite channel models such as fading models accounting for frequency-dependent atmospheric effects and for the case of mobile (legitimate) users.

\end{document}